# Optimization in the Loop: Implementing and Testing Scheduling Algorithms with SimuLTE

## Tutorial

Antonio Virdis

Dipartimento di Ingegneria dell'Informazione
University of Pisa
Pisa, Italy
a.virdis@iet.unipi.it

*Abstract*—**One of the main purposes of discrete event simulators such as OMNeT++ is to test new algorithms or protocols in realistic environments. These often need to be benchmarked against optimal/theoretical results obtained by running commercial optimization solvers. The usual way to do this is to have the simulator run in a standalone mode and generate (few) *snapshots*, which are then fed to the optimization solvers. This allows one to compare the optimal and suboptimal solutions *in the snapshots*, but does not allow to assess how the system being studied would *evolve over time* if the optimal solution was enforced every time. This requires optimization software to run directly in the loop of the simulation, exchanging information with the latter. The goal of this tutorial is to show how to integrate a commercial solver (CPLEX) into the simulation loop of the OMNeT++ environment. For this purpose, we propose two methods: a first one that uses a solver as an external program, and a second one that exploits a C-written API for CPLEX known as *Callable Library*. We then exemplify how to apply these two methods to SimuLTE, a simulation model for LTE cellular networks, implementing and testing a simple solution to a well-known resource-scheduling problem.**

## I. INTRODUCTION

Event-driven simulators such as OMNeT++ are widely used by researchers to test and evaluate complex communication systems. They generally aim at either evaluating the performance of a known system or testing new algorithms on the system itself. This becomes even more useful when a new algorithm has to be compared against other solutions in order to evaluate a performance improvement. It is a common approach in this case to use optimal/theoretical results as terms of comparison: the problem is formulated as a mathematical model and the optimal solution is obtained using commercial solvers. The obtained results are then used to assess how close the new algorithm is to the theoretical best value, for example by computing the optimality ratio of some metrics.

In order to carry out the above method, we need a means to feed the optimization problem with actual inputs from the simulated system. This is usually done by running simulation campaigns using the algorithm that has to be tested and generating *snapshots* of the system status, i.e. collections of all the variables of the system that are relevant to the problem being considered. Each snapshot is then used to create an instance of the optimization problem that is then solved *offline*, using a commercial solver as a standalone program. As we can see in the left part of Fig. 1, the evolution of the simulated system depends only on the results produced by the heuristic algorithm under test, whereas the output of the solver is used only for performance comparison. In other words the optimal solution generated by the solver is never fed back to the simulated system, thus the solver is actually *outside* the simulation loop. This method requires only few modification to the simulation, and is effective when evaluating single snapshots. In general it does not show the long-term effects of using the optimal solutions as the system evolves.

A possible alternative to this method is shown in the right part of Fig. 1: two instances of the system are created with the same initial conditions. The two instances evolve independently, one taking feedbacks only from the heuristic, the other one only from the solver. The results of both systems are finally compared, allowing us to analyze the steady-state performance of the system.

The purpose of this tutorial is to describe methods to integrate optimization solvers into the simulation loop. We will propose two techniques and highlight their pros and cons. Then we will apply these methods to an INET-based simulation module for LTE networks, namely SimuLTE [1][2], using a well-known problem as an example. In the rest of this paper, we will first describe the outline of the tutorial in Section II, then conclusions will be draw in Section III.

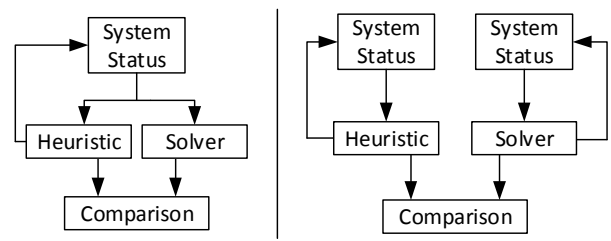

Fig. 1. Two examples of optimization usage in simulation: without feedback (left) and in the loop (right)

## II. Outline of the Tutorial

The tutorial will be divided into two parts, one detailing the two integration methods, and the other exemplifying them with a practical example. A brief description of these parts follows.

### A. Integrating optimization in the loop

First of all we describe two methods for integrating optimization into the simulation loop. We will use CPLEX [3] as an example of optimization software. The latter is a commercial product developed by IBM, made available to researchers via the IBM Academic Initiative. With reference to Fig. 2, we will consider a system composed of a Simulator (e.g. one based on OMNeT++) and an Optimization Tool. The former can be further split logically into one part implementing the system layers and another where the algorithm under study operates. The communication between the simulator and the Optimization Tool takes place whenever the algorithm is run and can be performed in four steps:

1. The allocation module reads all the relevant info from the system layers;

2. The allocation module builds an instance of the optimization problem and sends it to the optimization tool;

3. The optimization tool solves the problem and sends the solution back to the allocation module;

4. The allocation module parses the solution and enforces the algorithm decisions.

From a higher point of view we can say that steps 1 and 2 are responsible for feeding the optimization tool with the status of the system, whereas steps 3 and 4 implements the feedback operation.

The two integration methods we propose share the same general structure, but differ in the way the communication between the allocation module (i.e. the simulation environment) and the optimization tool is realized. The main idea behind the first method is to use CPLEX as an external program that is called during the simulation process. The tool being considered in this case is the CPLEX *Interactive Optimizer*, a command line interface to CPLEX. The communication between the two entities is made via file exchange, using a problem file in the LP format for step 2, and a solution file in XML format for step 3.

The second method we propose uses instead the CPLEX *Callable Library*. The latter is a set of C-written APIs that can be used to call CPLEX functions directly from the simulator code. For this purpose we first need to include the callable library into the OMNeT++ project. This way we can implement the communication in step 2 and 3 using simple C function calls. Moreover we will present a custom C++ interface for the *Callable Library* that can be used to simplify the writing of the problem.

We will discuss the pros and cons of the above methods, focusing on the trade-off between complexity of usage and performance (i.e. execution time). On one hand using the first method requires one only to write a file in LP format, which is rather simple, and to know how to parse an XML file.

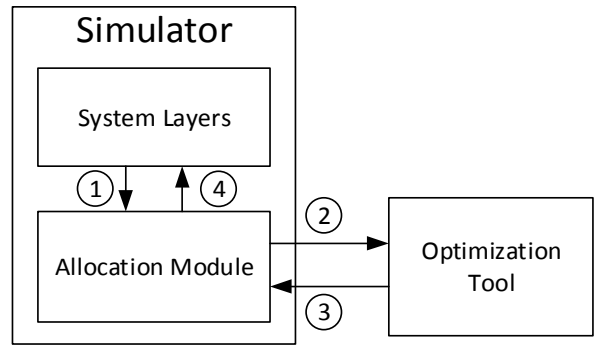

Fig. 2. Connection between a simulation module and an optimization tool

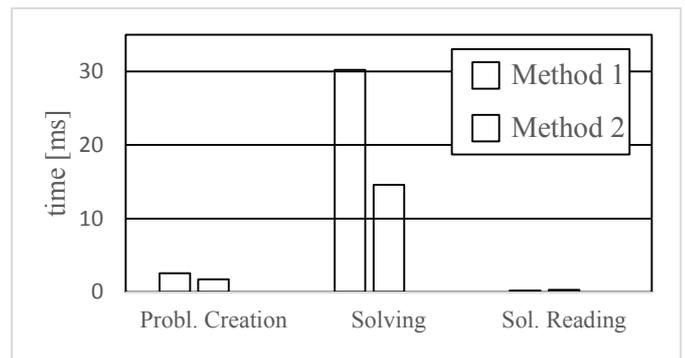

Fig. 3. Example of performance analysis of the two integration methods.

The second method requires instead knowledge of the Callable Library and its data structures. However, we will try to reduce the impact of this second issue with a custom C++ interface, as stated above. On the other hand, the first method is less efficient from the perspective of execution time. In fact it entails calling an external program, which requires frequent context switches, i.e. time consuming operations. The Callable Library is instead executed into the same context with a significant performance improvement. We will show how to measure the execution time of the two above methods, and how to evaluate the impact of each step of the process.

### B. Scheduling algorithms with SimuLTE

In the second part of the tutorial we apply the two above methods to a practical problem using the *multiband scheduling problem* (MBS) as an example. The latter is a resource allocation problem typical of OFDM systems. We consider it in the in the context of a specific cellular technology, namely LTE. For this purpose we will use SimuLTE, an INET-based simulation model, focusing on its resource scheduling capabilities. We start by briefly describing the LTE system, with particular emphasis on resource allocation. For each characteristic of the system we give an explanation of the modeling assumption of the corresponding element in SimuLTE.

We then discuss the idea behind the MBS problem and give detail on how we can solve it both at optimality and heuristically. We show how to create the optimization problem and how to implement both integration methods into SimuLTE. For this purpose we divide this process into three phases that are common to both methods:

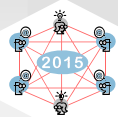

- a *problem creation* phase where the optimization problem is built;

- a *solving* phase during which the optimization tool is computing the solution;

- and a *solution reading* phase where the solution is read and stored into the simulator environment.

Finally we analyze the performance of the two integration methods from the point of view of execution time. This will help us to highlight the differences between them in order to better understand their pros and cons. An example of this analysis is shown in Fig. 3 where the execution time of each phase is compared.

## III. Conclusions

The goal of this tutorial is to analyze the process of integrating optimization solvers into the simulation loop of OMNeT++. The tutorial is organized in two parts. We start by proposing two integration methods: a first one that uses CPLEX as an external program and a second one that calls it directly within the simulator code. We analyze pros and cons of both methods, giving particular emphasis. Then in the second part we consider a practical example describing and solving the problem of multiband scheduling in the context of LTE networks. In order to simulate the latter we use SimuLTE, an INET based module, and we show how to implement the presented methods into it.